\documentclass[iop]{emulateapj}

\usepackage{epstopdf}
\usepackage{amsmath}

\newcommand{\Ha}{\ensuremath{\mathrm{H} \alpha}}
\newcommand{\Hb}{\ensuremath{\mathrm{H} \beta}}
\newcommand{\vlsr}{\ensuremath{v_{\mathrm{LSR}}}}

\newcommand{\sii}{\ensuremath{\textrm{[\ion{S}{2}]}}}
\newcommand{\nii}{\ensuremath{\textrm{[\ion{N}{2}]}}}
\newcommand{\Aliii}{\ensuremath{\textrm{\ion{Al}{3}}}}
\newcommand{\oiii}{\ensuremath{\textrm{[\ion{O}{3}]}}}
\newcommand{\oii}{\ensuremath{\textrm{[\ion{O}{2}]}}}
\newcommand{\oi}{\ensuremath{\textrm{[\ion{O}{1}]}}}
\newcommand{\Hei}{\ensuremath{\textrm{\ion{He}{1}}}}

\newcommand{\R}{\ensuremath{\, \mathrm{R}}}
\newcommand{\Hp}{\ensuremath{\mathrm{H}^+}}

\newcommand{\HHa}{\ensuremath{H_{n_e^2}}}
\newcommand{\Hn}{\ensuremath{H_{n_e}}}
\newcommand{\Hsii}{\ensuremath{H_{\sii}}}

\newcommand{\IHa}{\ensuremath{I_{\Ha}}}
\newcommand{\IHac}{\ensuremath{I_{\Ha}^c}}
\newcommand{\Isii}{\ensuremath{I_{\sii}}}

\newcommand{\cucm}{\ensuremath{\textrm{ cm}^{-3}}}

\newcommand{\kms}{\ensuremath{\textrm{ km s}^{-1}}}
\newcommand{\kpc}{\ensuremath{\, \mathrm{kpc}}}
\newcommand{\pc}{\ensuremath{\textrm{ pc}}}
\newcommand{\K}{\ensuremath{\textrm{ K}}}

\begin{document}


\title{\Ha\ and \sii\ emission from warm ionized gas in the Scutum-Centaurus Arm}
\submitted{Accepted by ApJ on April 7, 2014}
\author{Alex S. Hill\altaffilmark{1}, Robert A. Benjamin\altaffilmark{2}, L. Matthew Haffner\altaffilmark{3,4}, Martin Gostisha\altaffilmark{2}, \& Kathleen A. Barger\altaffilmark{5,6}}
\altaffiltext{1}{CSIRO Astronomy \& Space Science, Marsfield, NSW, Australia; alex.hill@csiro.au}
\altaffiltext{2}{Department of Physics, University of Wisconsin-Whitewater, Whitewater, WI, USA}
\altaffiltext{3}{Department of Astronomy, University of Wisconsin-Madison, Madison, WI, USA}
\altaffiltext{4}{Space Science Institute, Boulder, CO, USA}
\altaffiltext{5}{Department of Physics, University of Notre Dame, South Bend, IN, USA}
\altaffiltext{6}{NSF Astronomy \& Astrophysics Postdoctoral Fellow}

\begin{abstract}
We present Wisconsin H-Alpha Mapper \sii $\lambda 6716$ and \Ha\ spectroscopic maps of the warm ionized medium (WIM) in the Scutum-Centaurus Arm at Galactic longitudes $310^\circ < l < 345^\circ$. Using extinction-corrected \Ha\ intensities (\IHac), we measure an exponential scale height of electron density-squared in the arm of $\HHa = 0.30 \kpc$ (assuming a distance of $3.5 \kpc$), intermediate between that observed in the inner Galaxy and in the Perseus Arm. The $\sii / \Ha$ line ratio is enhanced at large $|z|$ and in sightlines with faint \IHac. We find that the $\sii / \Ha$ line ratio has a power law relationship with \IHac\ from a value of $\approx 1.0$ at $\IHac < 0.2 \R$ (Rayleighs) to a value of $\approx 0.08$ at $\IHac \gtrsim 100 \R$. The line ratio is better correlated with \Ha\ intensity than with height above the plane, indicating that the physical conditions within the WIM vary systematically with electron density. We argue that the variation of the line ratio with height is a consequence of the decrease of electron density with height. Our results reinforce the well-established picture in which the diffuse \Ha\ emission is due primarily to emission from {\em in situ} photoionized gas, with scattered light only a minor contributor.
\end{abstract}

\keywords{dust, extinction --- ISM: atoms --- ISM: kinematics and dynamics --- ISM: structure}
\maketitle

\section{Introduction}

The warm ionized medium (WIM) traces energy transport in the diffuse interstellar medium (ISM) of star-forming galaxies through supernova-driven turbulence \citep{Armstrong:1995ho,Haverkorn:2008kt,Chepurnov:2010bt,Hill:2008gv,deAvillez:2012hp} and photoionization. Because of its large ($\sim 1 \kpc$; \citealt{Haffner:1999hi,Gaensler:2008bq,Savage:2009kj}) scale height, its weight is substantial, making it an important contributor to the hydrostatic pressure of the ISM \citep{Boulares:1990ba}.
Evidence for the existence of the WIM comes from an absorption signature in the synchrotron spectrum \citep{Hoyle:1963ve}, pulsar dispersion \citep{Manchester:1981ge,Taylor:1993il}, Faraday rotation \citep{Mao:2010eg,Foster:2013gg}, and faint emission \citep{Reynolds:1973jo,Dettmar:1990tv} and absorption \citep{Savage:2009kj,Howk:2012gt} lines.

Classical \ion{H}{2} regions, which are ionized by a local association of hot stars, are distinct from the WIM: the \ion{H}{2} regions have different spectral signatures \citep{Madsen:2006fw}, a much lower scale height ($\sim 50 \pc$ in the Milky Way), and a higher dust content \citep{Kreckel:2013fh,Rueff:2013ep}. However, like \ion{H}{2} regions, the WIM is associated with star formation. It is found in every star-forming galaxy, accounting for $59 \pm 19\%$ of the \Ha\ flux in normal star-forming galaxies \citep{Oey:2007cj} and $\gtrsim 90 \%$ of the \Hp\ mass \citep{Haffner:2009ev}. The WIM is primarily ionized by photons which escape from \ion{H}{2} regions and then travel through low-density pathways around neutral hydrogen established by stellar feedback and turbulence \citep{Reynolds:1990kw,Ciardi:2002bz,Wood:2010dg}.

The \Ha\ emission line provides the bulk of the information about the distribution of the WIM, with \Ha\ from the WIM of the Milky Way detected in every direction \citep{Haffner:2003fe,Haffner:2009ev}. Up to $\approx 20 \%$ of the faint \Ha\ flux may be scattered light which originated in \ion{H}{2} regions, with the WIM contributing the remainder \citep{Reynolds:1973jo,Wood:1999bd,Witt:2010gd,Brandt:2012fw}.
The relative intensities of collisionally-excited optical emission lines, primarily \nii $\lambda 6584$ and \sii $\lambda 6713$ (hereafter \nii\ and \sii), provide the best probes of the ionization state and temperature of the gas in the WIM. In general, the line intensity ratios $\sii/\Ha$ and $\nii/\Ha$ track each other and increase with decreasing \Ha\ intensity both in the Milky Way \citep{Haffner:1999hi,Madsen:2006fw} and in other galaxies; infrared and ultraviolet collisionally-excited lines trace similar physical trends \citep{Rand:2008dj,Rand:2011hd,Howk:2012gt}. Also, \oii\ is relatively bright while \oi, \oiii, and \Hei\ are faint in the WIM \citep{Domgorgen:1994kk,Reynolds:1995ki,Reynolds:1998ji,Mierkiewicz:2006ky}. In combination, these observations indicate that the WIM has a higher temperature ($8000 \K$ compared to $6000 \K$) but lower ionization state (O$^+$ and S$^+$, not O$^{++}$ and S$^{++}$, are the dominant ions) than classical \ion{H}{2} regions. 

The physical cause of the trends in temperature and ionization state inferred from the observed line ratios must relate to the heating and ionization of the gas. One would expect the radiation field to harden as it escapes \ion{H}{2} regions, but this would produce the opposite of the observed line ratio trends: a harder ionizing radiation field would produce higher ionization states in the WIM. Models in which the radiation field is dilute ($U \sim 10^{-3} - 10^{-4}$ photons per electron) --- allowing ions time to recombine before encountering an ionizing photon --- can explain the low ionization state \citep{Mathis:1986et}, but variations in $U$ cannot explain the observed constancy of $\sii/\nii$. A supplemental heating source which scales less steeply with density than photoionization heating (and therefore dominates at densities $n_e \lesssim 0.1 \cucm$) appears necessary to explain the inferred temperature trend \citep{Reynolds:1999jp}. In edge-on star-forming galaxies, a clear relationship between line ratios and height is observed \citep{Domgoergen:1997wh,Tullmann:2000td,Otte:2002dr,Hoopes:2003ec}. However, because the \Ha\ intensity also decreases with height, it is unclear whether the line ratios are fundamentally a function of height, density, or a combination of the two.

In this paper, we use Wisconsin H-Alpha Mapper (WHAM) observations to investigate the physical properties of the WIM associated with the Scutum-Centaurus Arm, seen edge-on in the Galactic longitude range $310^\circ \lesssim l \lesssim 340^\circ$ in the local standard of rest velocity range $-80 \kms < \vlsr < -40 \kms$ \citep[e.~g.][]{Benjamin:2008uh}, analogous to the work done in the Perseus Arm in the outer Galaxy by \citet{Haffner:1999hi} and \citet{Madsen:2006fw}. We describe our observations in Section~\ref{sec:obs} and present them in Section~\ref{sec:maps}. We estimate an extinction correction from \ion{H}{1} data, tested against \Hb\ spectra in a few sightlines, in Section~\ref{sec:dust}. In Section~\ref{sec:height}, we measure the scale height of the WIM in the Scutum-Centaurus Arm. We present $\sii/\Ha$ line ratios and interpret them as a measurement of the temperature of the gas in Section~\ref{sec:lineratios}, addressing the degeneracy between height and density as underlying causes of the observed line ratio trends. We present conclusions in Section~\ref{sec:discussion}, arguing that scattered light is a minor contributor to the \Ha\ light attributed to the WIM \citep[in contrast to the model of][]{Seon:2012ci}. Finally, we summarize the paper in Section~\ref{sec:summary}.

\section{Observations} \label{sec:obs}

Our data were obtained with WHAM, a Fabry-Perot spectrometer designed to detect faint optical emission lines from the WIM. The instrument, described in detail by \citet{Haffner:2003fe}, has been located at the Cerro Tololo Inter-American Observatory in Chile since 2009 after twelve years at the Kitt Peak National Observatory. WHAM has a $12.5 \kms$ spectral resolution and a $1^\circ$ beam; all spatial information within the beam is lost in the observing mode used here. We use preliminary \Ha\ data from the WHAM Southern Sky Survey \citep{Haffner:2010tha} and preliminary \sii\ data from an \sii\ survey of the southern Galactic plane ($|b| \lesssim 30^\circ$; \citealt{Gostisha:2013ww}).
The \Ha\ and \sii\ data were each obtained with the block mapping technique described by \citet{Haffner:2003fe} with the $200 \kms$-wide velocity windows chosen to include emission at $\vlsr \approx 0$ as well as to the terminal velocity at up to $\vlsr \sim -120 \kms$. In some blocks, particularly near the Galactic Center where the terminal velocity is large, we obtained a second \sii\ spectrum with a shifted velocity window to capture all velocities in which emission is expected and combined the two spectra.

The \Ha\ observations, obtained for the WHAM Southern Sky Survey, consist of $30$~s exposures, reaching a $3\sigma$ sensitivity of $\approx 0.15 \R$. We have applied a flat field, then fit and subtracted a Gaussian term corresponding to the geocoronal \Ha\ emission as well as a constant background term. At the sensitivity limit of WHAM, atmospheric emission lines of $\sim 0.03-0.15 \R$ dominate the background in the \Ha\ and \sii\ spectral windows \citep{Hausen:2002iu}. We have not fit an atmospheric template to our \Ha\ data, so our effective sensitivity is determined by the atmospheric emission lines. We measured the intensity of the \ion{H}{2} region surrounding $\lambda$~Orionis from Kitt Peak and Cerro Tololo to calibrate for changes to the instrumental throughput. The intensity calibration assumes that the North American Nebula has an \Ha\ intensity of $850 \R$ within a $50 \arcmin$ beam \citep{Scherb:1981dl,Haffner:2003fe}.

We have obtained \sii\ spectra with 60~s exposures for much of the southern Galactic plane \citep{Gostisha:2013ww}. For these data, we applied a flat field then fit an atmospheric template and constant baseline to achieve the noise-limited $3 \sigma$ sensitivity of $\approx 0.1 \R$.

We also obtained deep (240~s) \Hb\ spectra in three directions and for one $\approx 7^\circ \times 7^\circ$ block (centered near $(l,b)=(332^\circ,-2^\circ)$) with 60~s exposures. We have applied a flat field and subtracted the geocoronal \Hb\ emission, an atmospheric template, and a linear polynomial from the \Hb\ data. Following \citet{Madsen:2005ft}, we calibrated the relative intensities of the \Ha\ and \Hb\ spectra against the WHAM calibration sightline in the \ion{H}{2} region surrounding Spica ($\alpha$~Vir); we assume this sightline has low extinction due to its small distance ($80 \pc$), high latitude ($b=+50^\circ$), and the low observed $E(B-V) = +0.01$ to the star \citep{Galazutdinov:2008hj}.

\section{Maps and spectra} \label{sec:maps}

\begin{figure}
\plotone{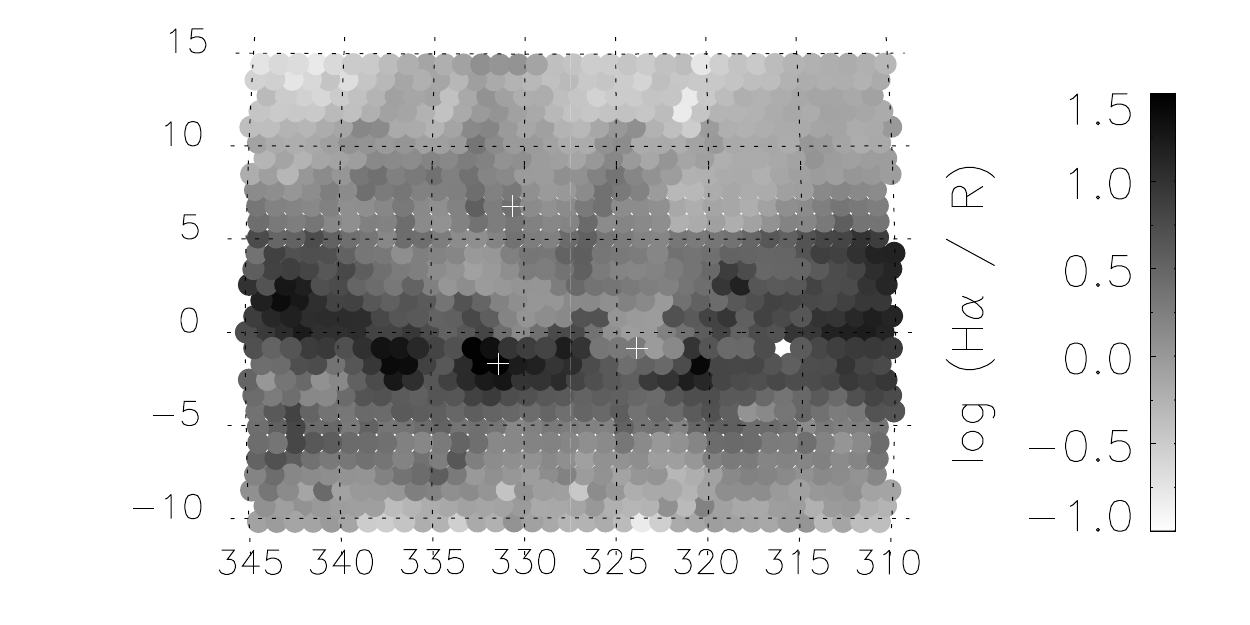}
\plotone{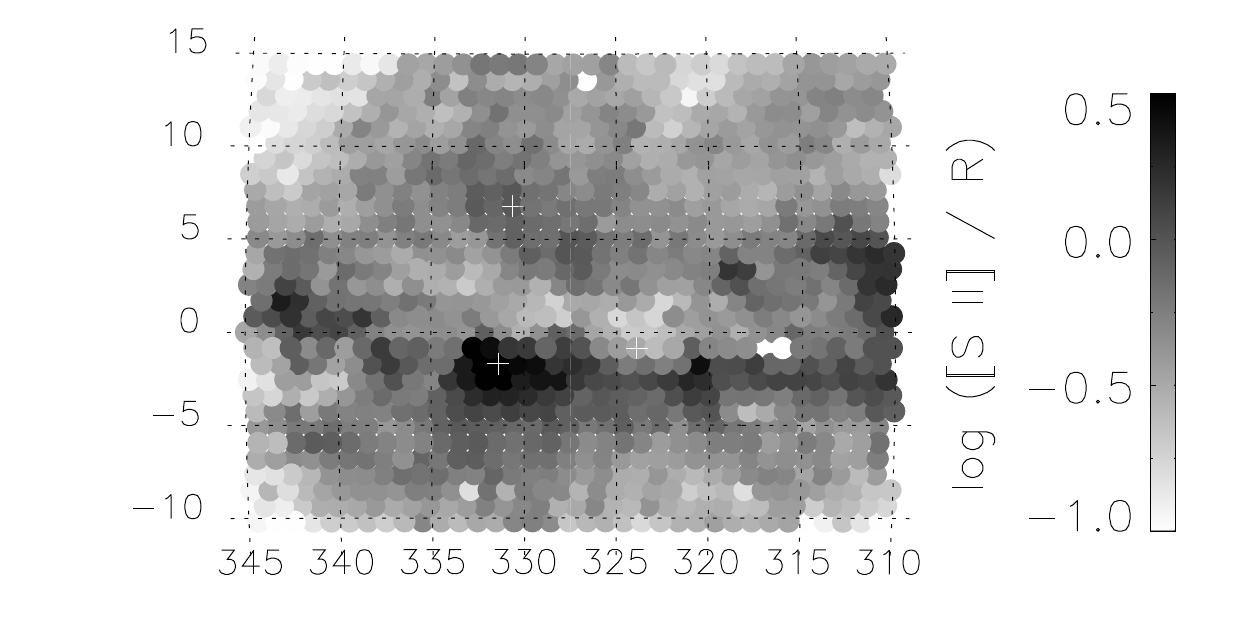}
\plotone{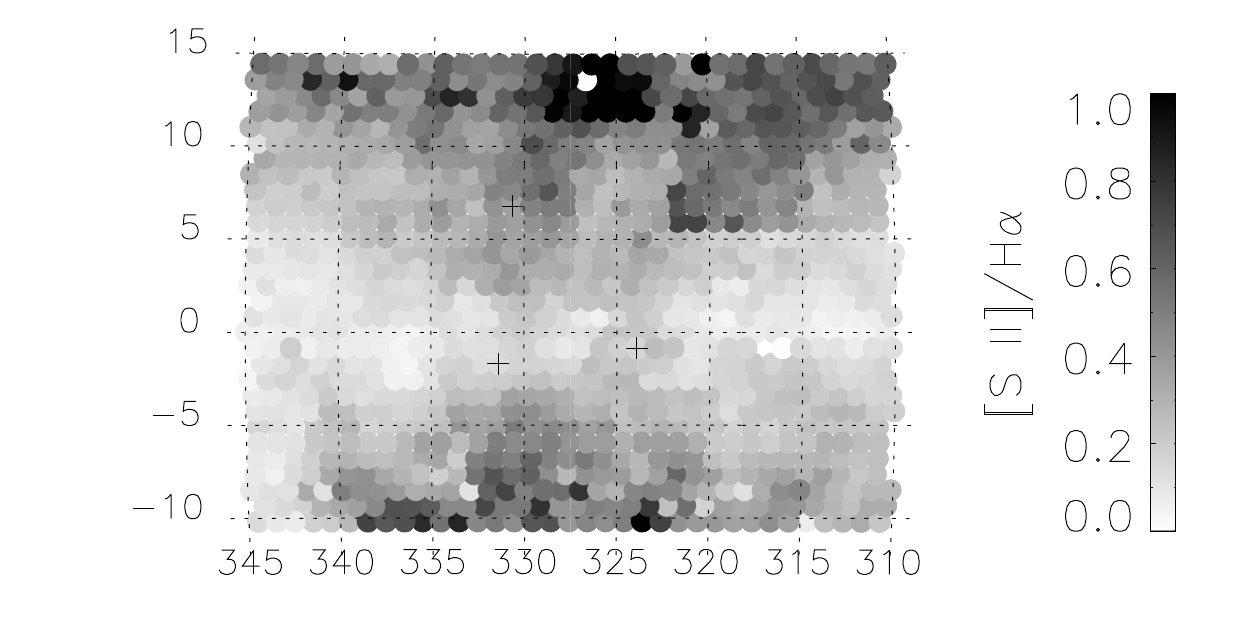}
\caption{Maps of \IHa, \Isii, and the line ratio $\sii/\Ha$ in Galactic coordinates, integrated over $-80 \kms < \vlsr < -40 \kms$.}
\label{fig:maps}
\end{figure}

\begin{figure}[tb]
\plotone{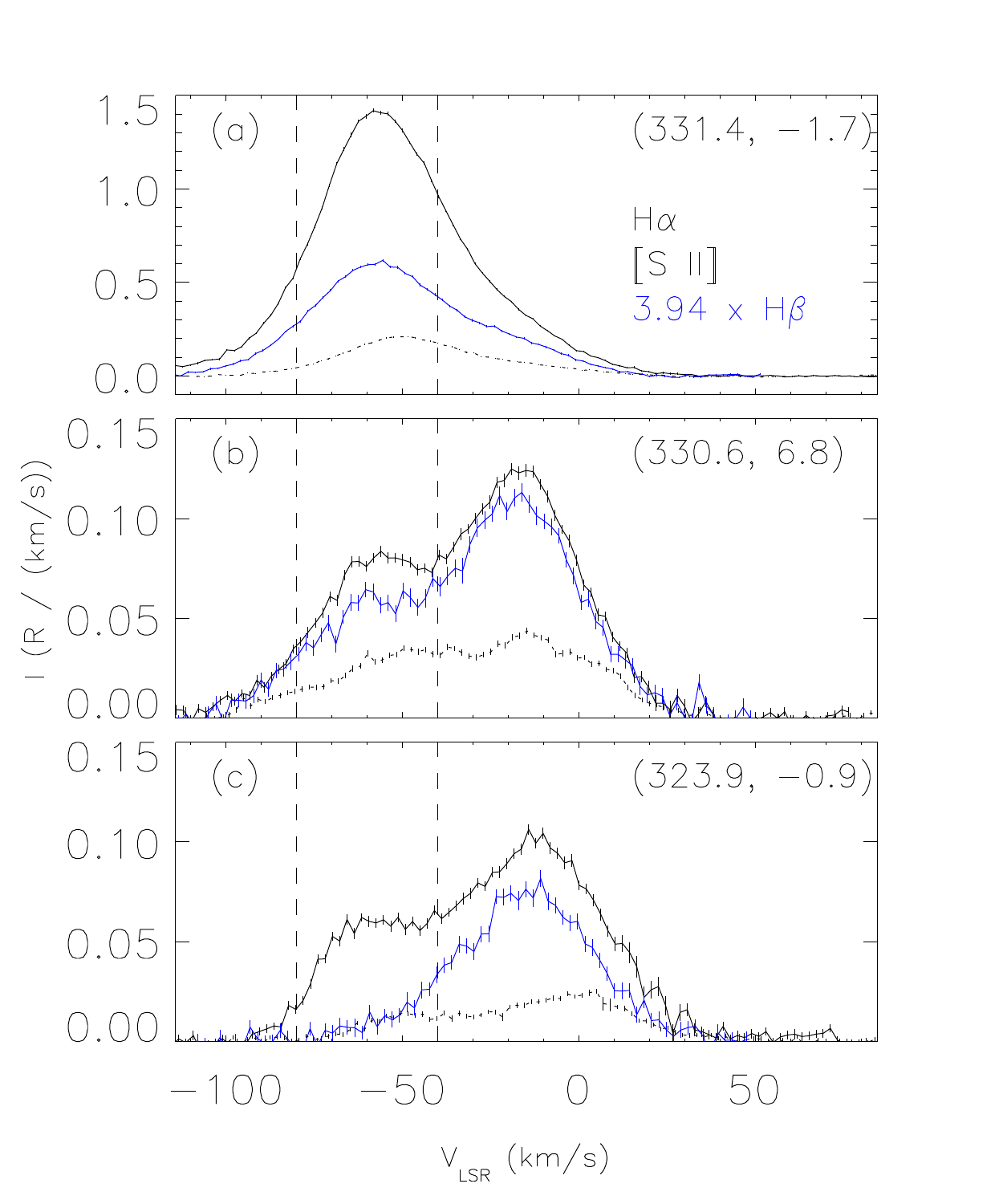}
\caption{\Ha\ (solid lines), \sii\ (dotted lines), and \Hb\ (blue lines) spectra towards representative sightlines at the indicated latitude and longitude and marked with $+$ signs in Figure~\ref{fig:maps}. The vertical dashed lines denote the velocity range integrated over in Figures~\ref{fig:maps} and \ref{fig:extinction}$-$\ref{fig:ratio}. The \Hb\ spectra have been multiplied by $3.94$; with no extinction and assuming photoionization, these spectra would have the same intensity as the \Ha\ spectra.}
\label{fig:spectra}
\end{figure}

We show WHAM maps of \IHa, \Isii, and the line ratio $\sii/\Ha$ from the Scutum-Centaurus Arm in Figure~\ref{fig:maps}. To isolate the emission, we used a \citet{Clemens:1985dp} rotation curve with spiral arm positions from \citet{Taylor:1993il}, compiled by \citet{Benjamin:2008uh,Benjamin:2009ks}. From this rotation curve, we used emission integrated over $-80 \kms < \vlsr < -40 \kms$ to represent the arm. The midplane is brightest, with a decrease in intensity with $|b|$ evident in both \Ha\ and \sii. The midplane \Ha\ emission is faint over $321^\circ \lesssim l \lesssim 326^\circ$; no \ion{H}{2} regions are evident in this region from $5$~GHz radio continuum emission \citep{Haynes:1978uw}. There is a region of bright emission centered at $(l,b) = (332^\circ,-2^\circ)$. The $\sii/\Ha$ line ratio is typically $\approx 0.1$ in the midplane, but is higher at higher latitudes and in $321^\circ \lesssim l \lesssim 326^\circ$.

In Figure~\ref{fig:spectra}, we show spectra for three sightlines. These spectra have two primary components in both \Ha\ and \sii. The component which peaks near $\vlsr = -15 \kms$ is the foreground Sagittarius-Carina Arm; the component which peaks near $\vlsr = -60 \kms$ is the Scutum-Centaurus arm. There is an inflection point near $\vlsr = -40 \kms$ in these sightlines; at $\vlsr < -40 \kms$, the emission in both lines is primarily from the Scutum-Centaurus Arm, justifying our choice to integrate over $-80 \kms < \vlsr < -40 \kms$ to extract emission from Scutum-Centaurus.

\section{Dust extinction} \label{sec:dust}

\begin{figure}[tb]
\plotone{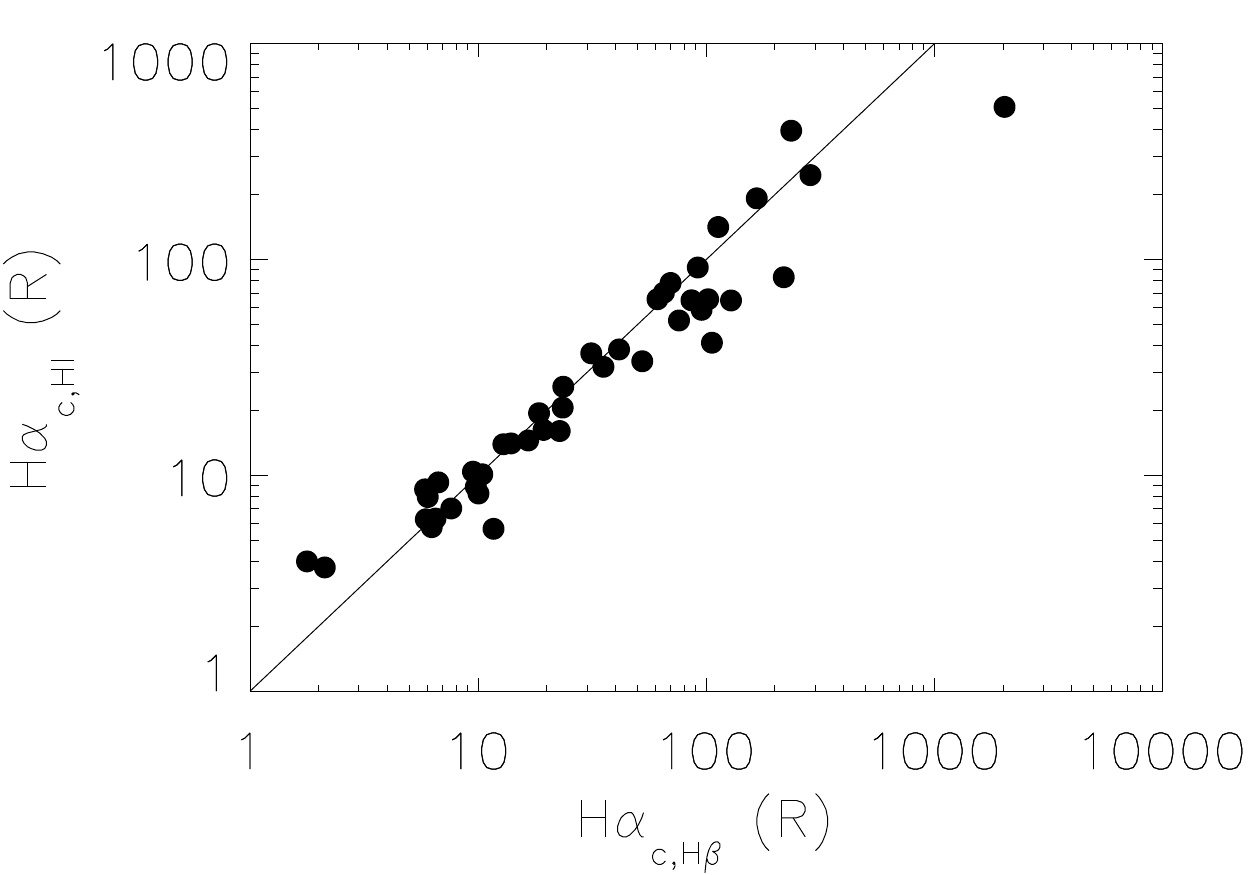}
\caption{Extinction-corrected \Ha\ derived from \ion{H}{1} data (vertical axis) versus the extinction-corrected \Ha\ derived from the observed $\Ha/\Hb$ line ratio (horizontal axis) from the region $328\arcdeg < l < 336 \arcdeg$, $-5\arcdeg < b < 0\arcdeg$. A unity line is drawn to guide the eye. Data are integrated over $-80 \kms < \vlsr < -40 \kms$.}
\label{fig:extinction}
\end{figure}

The three directions in Figure~\ref{fig:spectra} probe sightlines with varying extinction. Panel $a$ shows data in the midplane in a region of bright emission; panel $b$ shows $7\arcdeg$ above the midplane near the longitude of panel $a$; and panel $c$ shows the midplane in a region with fainter \Ha\ emission than the surroundings. We use the \Hb\ spectra to test for extinction, following \citet[and references therein]{Madsen:2005ft}. In the absence of extinction, case B recombination requires that the $\Ha/\Hb$ line ratio in $\sim 10^4 \K$ photoionized gas is $3.94$ in the photon units used in this paper \citep{Osterbrock:1989ul}. For emission from the Scutum-Centaurus Arm, we find $\Ha/\Hb = 8.88 \pm 0.16, 5.1 \pm 0.5,$ and $19 \pm 9$ for the three sightlines respectively, corresponding to $A(V) = 2.54 \pm 0.06, 0.8 \pm 0.3,$ and $4.9 \pm 1.2$ magnitudes more extinction than towards the Spica \ion{H}{2} region calibration direction.

Because we do not have a wide-area \Hb\ map for this region, we apply an estimated correction for extinction to the \Ha\ intensities based on the foreground dust column inferred from the \ion{H}{1}. We use the \ion{H}{1} column from GASS \citep{Kalberla:2010jo} integrated over $-40 \kms < \vlsr < +20 \kms$ combined with the empirical relationship between \ion{H}{1} column density and color excess from \citet{Bohlin:1978dw} and the extinction curves from \citet{Cardelli:1989dp}, assuming $R_V = 3.1$.

We confirmed the effectiveness of this technique for our data by comparing the extinction-corrected \Ha\ intensities, \IHac, to corrected intensities measured using the Balmer decrement, ${\IHac}_{,\Hb}$. We obtained \Hb\ spectra with WHAM for $N=42$ sightlines (one WHAM block) including the sightline shown in Figure~\ref{fig:spectra}$a$, $(331.4^\circ, -1.7^\circ)$. For each of these sightlines, we used the \Hb\ intensity to calculate  ${\IHac}_{,\Hb}$ following \citet{Madsen:2005ft}. The extinction-corrected \Ha\ intensities derived with these two methods are compared in Figure~\ref{fig:extinction}. The median ratio ${\IHac}_{,\mathrm{HI}}/{\IHac}_{,\Hb}$ is $0.97$ with a median absolute deviation of $0.15$, which we incorporate as the fractional uncertainty introduced by the extinction correction.

\section{Scale height of the ionized gas} \label{sec:height}

\begin{figure}
\plotone{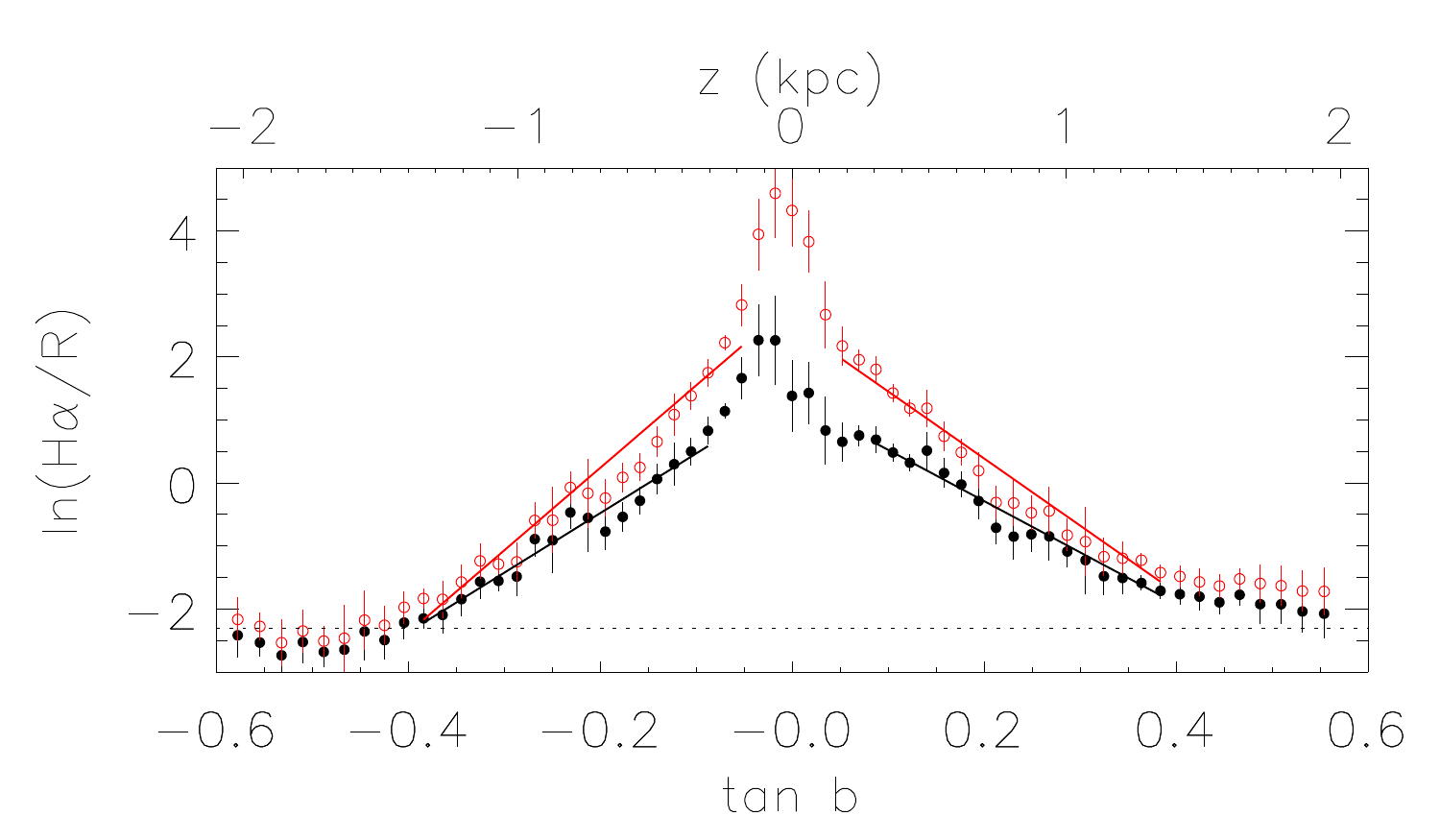}
\caption{\Ha\ intensity (filled) and extinction-corrected \IHac\ (open, red circles) as a function of Galactic latitude. The lines show fits to equation~(\ref{eq:height}). The values for each latitude are the median of the values in $1^\circ$ latitude bins in the longitude range $320^\circ < l < 340^\circ$, with the error bars showing the median absolute deviation about the median. The approximate instrumental sensitivity determined by atmospheric emission lines is shown with a horizontal dotted line. Heights $z$ assume $D=3.5 \kpc$ to the Scutum-Centaurus Arm.}
\label{fig:scale_height}
\end{figure}

\begin{deluxetable}{llrlrr}
\tabletypesize{\footnotesize}
\tablecolumns{6}
\tablewidth{0pt}
\tablecaption{Scale height measurements}
\tablehead{\colhead{$b_{\mathrm{min}}$} & \colhead{$b_{\mathrm{max}}$} & \colhead{$R_G$}	& \colhead{$\ln (I_0 / \R)$} & \colhead{$D/\HHa$} & \colhead{$\HHa$} \\
 & & \colhead{(kpc)} & & & \colhead{(pc)}}
\startdata
\cutinhead{Scutum-Centaurus Arm}
$  5^\circ$ & $ 21^\circ$ & $5.5$ & $1.34 \pm 0.12$ & $ 8.1 \pm 0.4$ & $430 \pm  50$ \\
$-21^\circ$ & $ -5^\circ$ & $5.5$ & $1.41 \pm 0.16$ & $ 9.5 \pm 0.6$ & $370 \pm  40$ \\
\cutinhead{Scutum-Centaurus Arm, extinction-corrected}
$  3^\circ$ & $ 21^\circ$ & $5.5$ & $2.52 \pm 0.10$ & $10.7 \pm 0.4$ & $330 \pm  40$ \\
$-21^\circ$ & $ -3^\circ$ & $5.5$ & $2.86 \pm 0.11$ & $13.1 \pm 0.5$ & $270 \pm  30$ \\
\cutinhead{Perseus Arm \citep{Haffner:1999hi}}
$-30^\circ$	& $-15^\circ$	& $10$	& $1.73 \pm 0.19$	& $4.91 \pm 0.39$	& $400 \pm 30$ \\
\cutinhead{Inner Galaxy, extinction-corrected \citep{Madsen:2005ft}}
$-5^\circ$ & $-0.8^\circ$		& $6.0$	& $4.9 \pm 0.5$	& $32^{+7}_{-10}$	& $95^{+30}_{-20}$ \\
$-5^\circ$ & $-0.8^\circ$		& $4.9$	& $5.5 \pm 0.5$	& $33^{+6}_{-11}$	& $135^{+45}_{-25}$ \\
$-5^\circ$ & $-0.8^\circ$		& $4.1$	& $6.3 \pm 0.9$	& $40^{+11}_{-24}$	& $150^{+90}_{-40}$
\enddata
\tablecomments{We have adjusted $\HHa$ for the Perseus Arm from the value reported by \citet{Haffner:1999hi} to account for the more recent measurement of $D = 1.95 \pm 0.04 \kpc$ towards $l=134\arcdeg$ \citep{Xu:2006it,Reid:2014td}. \citet{Madsen:2005ft} report $\Hn$ and $\phi^{1/2} n_e$; we derive their values and uncertainties for columns $4-6$ assuming no uncertainty in $D$. All scale heights do not account for the uncertainty in $D$, which is likely larger than the statistical uncertainties reported in the table.}
\label{tbl:h}
\end{deluxetable}

To estimate the scale height of free electrons in the Scutum-Centaurus Arm, we consider the \Ha\ intensity as a function of Galactic latitude, shown in Figure~\ref{fig:scale_height}.
We fit the data in Figure~\ref{fig:scale_height} to
\begin{equation} \label{eq:height}
\ln \IHa = \ln I_0 - \frac{D}{\HHa} \tan |b|,
\end{equation}
where $I_0$ is the midplane \Ha\ intensity and $D$ is the distance to the arm in the midplane. \citet{Reid:2014td} fit a logarithmic spiral with a pitch angle of $19.3\arcdeg$ to masers in the Scutum-Centaurus Arm in the range $6\arcdeg < l < 19\arcdeg$ with parallax distances. Assuming their fit applies through $l=0$ to the area we study, the distance to the arm ranges from $D=2.3 \kpc$ at $l=340\arcdeg$ to $D=2.7 \kpc$ at $l=320\arcdeg$. A Wolf-Rayet star thought to be in the arm at $l=353\arcdeg$ is at $D=3.5 \pm 0.3 \kpc$ \citep{Drew:2004kz}. Adjusting the \citet{Reid:2014td} results to use a pitch angle $\psi = 12\arcdeg$ (instead of their best fit, $\psi = 19.8\arcdeg \pm 2.6\arcdeg$) would make the distance to the arm at $320\arcdeg < l < 340\arcdeg$ roughly consistent with $3.5 \pm 0.3 \kpc$. There is additional uncertainty because the \Ha-emitting gas is not necessarily spatially coincident with the star-forming regions traced by masers. We thus consider the distance to be $3.5 \pm 0.3 \kpc$ and emphasize that we directly measure the quantity $D/\HHa$, not the scale height.

We fit equation~(\ref{eq:height}) using data above and below the plane separately. At $\tan |b| \gtrsim 0.4$, the \Ha\ intensities reach the noise level and thus are not used in the fit. The slope increases near the plane, likely due to classical \ion{H}{2} regions, which have a scale height of $\approx 50 \pc$ \citep{Gomez:2001fa}. Because this increase is only $\approx 3$ WHAM beams, we cannot reliably estimate the scale height of the \ion{H}{2} region component here. We thus also exclude $\tan |b| < 0.05$ from our fit. The results are listed in Table~\ref{tbl:h} fitting the \Ha\ data with and without the extinction correction. 

\label{sec:fillfrac}

We now use the measured $n_e^2$ scale height to estimate the scale height of the WIM.
Following previous authors \citep{Reynolds:1991ft,Heiles:1987wa,Berkhuijsen:2008kt,Gaensler:2008bq}, we adopt a simple model in which the electron density is $n_e = n_c(z)$ within the WIM and $n_e = 0$ elsewhere; the volume of the WIM is a fraction $f(z)$ of the total volume. Due to the approximately lognormal distribution of $n_e$ in the real WIM, this definition of $f(z)$ underestimates the true volume of the WIM by a factor of $\sim 2$ \citep{Hill:2008gv}, but we ignore this discrepancy here. With the assumption that $f(z)$ is constant with height \citep{Haffner:1999hi}, the scale height of the WIM in the Scutum-Centaurus arm is $\Hn = 2\HHa = 0.59 \pm 0.04 \kpc$. However, there is no a priori reason to expect $f(z)$ to be constant; given the smaller scale heights of other cold and warm phases of the ISM than that of the WIM, one would expect $f(z)$ to increase with height from the midplane to at least $|z| = 1 \kpc$ \citep[see Fig.~11 in the Erratum of][]{Hill:2012ic}. Therefore, following \citet{Gaensler:2008bq}, we assume
\begin{equation}
f(z) = f_0 e^{+|z|/H_f}
\end{equation}
and
\begin{equation}
n_c(z) = n_{c,0} e^{-|z|/H_{n,c}},
\end{equation}
defining $H_f$ and $H_{n,c}$ as the scale heights of the filling fraction and characteristic density and $f_0$ and $n_{c,0}$ as the midplane values. The scale heights are then \citep{Gaensler:2008bq}
\begin{equation}
H_{n,c} = \frac{\HHa \Hn}{\Hn - \HHa}
\end{equation}
and
\begin{equation}
H_f = \frac{\HHa \Hn}{\Hn - 2 \HHa}.
\end{equation}
If $\Hn = 2 \HHa$, then $H_f \rightarrow \infty$ and $H_{n,c} = \Hn$, recovering the uniform filling fraction result of \citet{Haffner:1999hi}.

\begin{figure*}
\plottwo{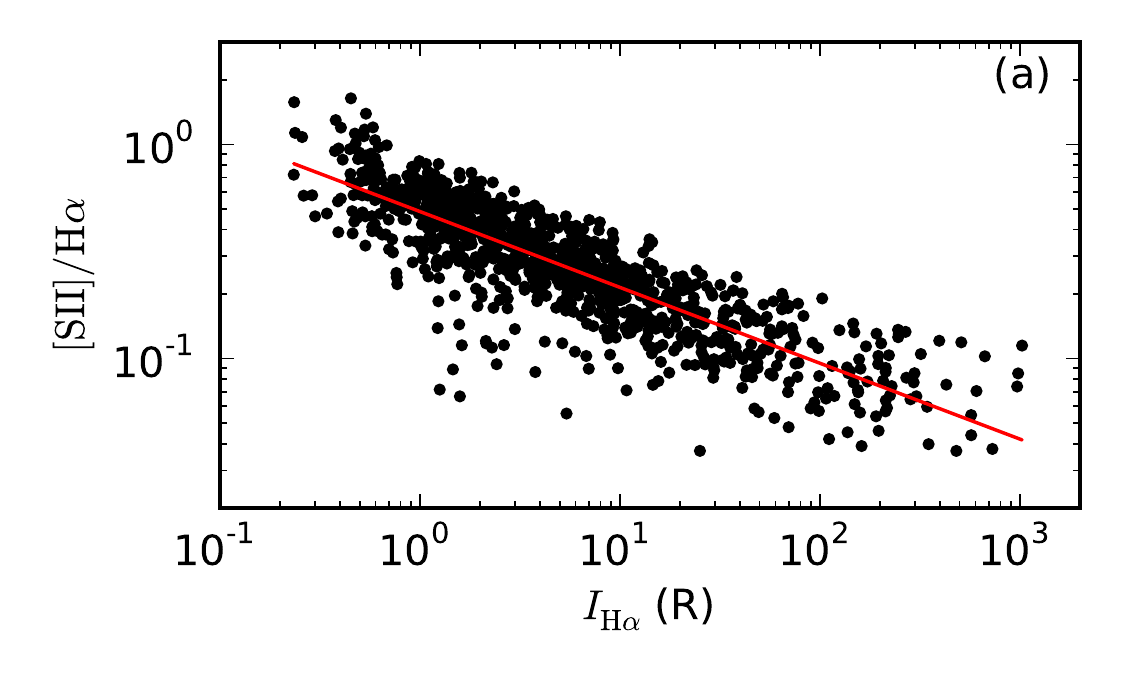}{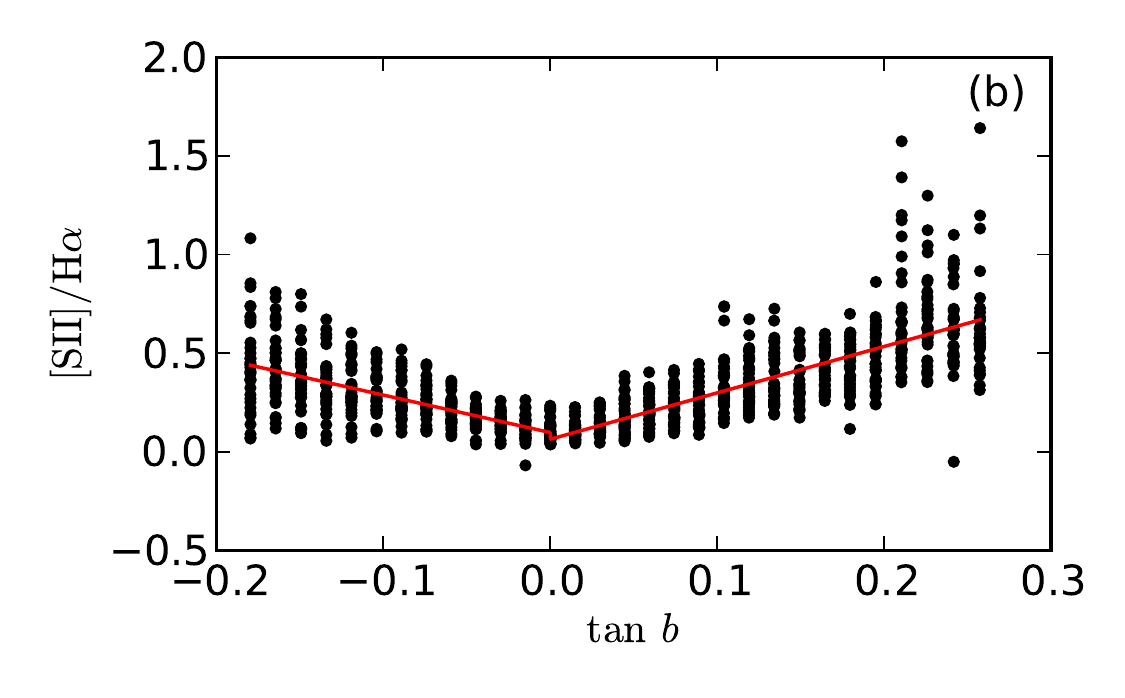} \\
\plottwo{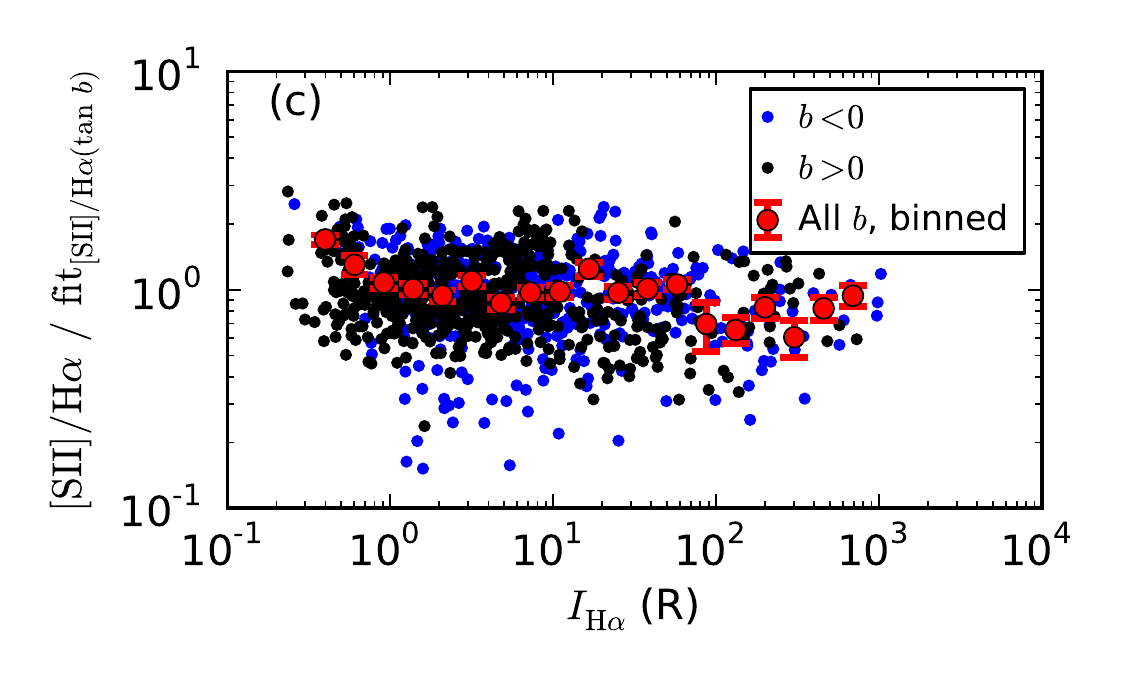}{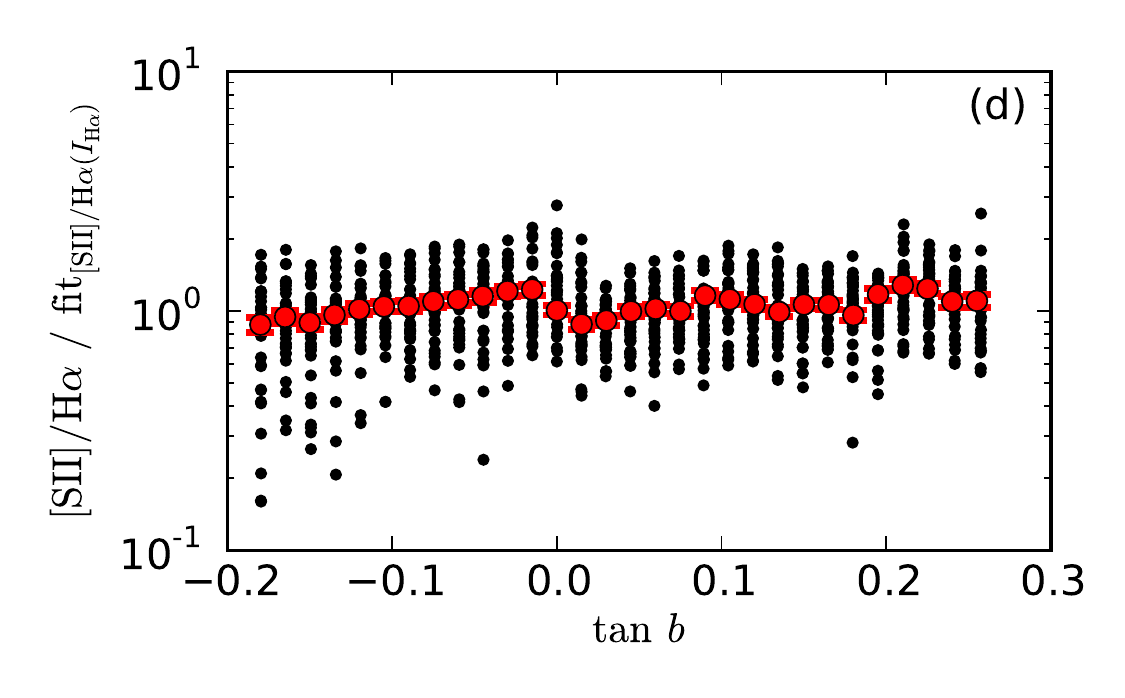}
\caption{Line ratio $\sii/\Ha$ as a function of \IHac\ (panel $a$) and $\tan b$ (panel $b$) for the data shown in Figure~\ref{fig:maps}. Red circles show the data within $0.1$~dex bins as described in the text. The solid line shows equation~(\ref{eq:ratio}) fit to the binned data. Panels $c$ and $d$ show these data divided by fits as described in Section~\ref{sec:lineratios}. Panel $c$ is analogous to panel $a$ with the height dependence (red line in panel $b$) divided out, while panel $d$ is analogous to panel $b$ with the \Ha\ intensity dependence (red line in panel $a$) divided out. Red points in panels $c$ and $d$ show the mean and standard deviation of the mean within bins.}
\label{fig:ratio}
\end{figure*}

The scale height inferred from pulsars in the interarm region near the Sun ($|b| > 40^\circ$) is $\Hn = 1.4 \pm 0.2 \kpc$ \citep{Savage:2009kj}. The \Aliii\ column towards stars --- primarily at low latitudes and including many sightlines through the Scutum-Centaurus Arm as well as many sightlines through interarm regions and other arms --- and towards extragalactic sources at all latitudes yields $\Hn = 0.90^{+0.62}_{-0.33} \kpc$ \citep{Savage:2009kj}. If we assume that $\Hn = 1.4 \kpc$ in the arm as it is locally, we find $H_f = 0.51 \kpc$ and $H_{n,c} = 0.37 \kpc$ in the arm. The data are consistent with a constant filling fraction with height only if $\Hn = 0.6 \kpc$, the $1 \sigma$ lower limit allowed by the \Aliii\ data.

The data in Figure~\ref{fig:scale_height} suggest the presence of a two-component vertical distribution of diffuse \Ha-emitting gas. We separately fit the extinction-corrected data in $5\arcdeg < |b| < 11\arcdeg$ and $11\arcdeg < |b| < 21\arcdeg$. At $|b| < 11\arcdeg$, we find $D/\HHa \approx 17.4$ (corresponding to $\HHa \approx 200 \pc$), while at $|b| > 11\arcdeg$, we find $D/\HHa \approx 8.5$ ($\HHa = 430 \pc$). We will report a further analysis of this behavior elsewhere.

\section{Line ratio and physical conditions} \label{sec:lineratios}

To diagnose the physical conditions in the ionized gas, we plot the $\sii/\Ha$ line ratio as a function of \Ha\ intensity and $\tan b$ (or height, $z=D \tan b$) in Figure~\ref{fig:ratio}. The line ratio is low, $\sii/\Ha \approx 0.1$, close to the midplane and in high \Ha\ intensity sightlines. The line ratio increases linearly with increasing $z$ and follows a power law with $\IHa$.

We fit the data in Figure~\ref{fig:ratio}$a$ to a power law,
\begin{equation} \label{eq:ratio}
\frac{\sii}{\Ha} \propto (\IHa)^{-a}.
\end{equation}
Because the uncertainties are systematically larger for data with a low $\sii/\Ha$ value, we bin the data, using the median and median absolute deviation from the median divided by $\sqrt{N}$ as the binned value and uncertainty, respectively. The fit parameter $a = 0.45 (0.33)$ for a weighted $\chi^2$ minimization fit of the uncorrected (extinction-corrected) data. The reduced chi-squared parameter for this fit is $\chi_r^2 = 35 (30)$. There is evidence for a flattening of the data at $\IHac \gtrsim 110 \R$, so we excluded these data from our fit.

We fit the data in Figure~\ref{fig:ratio}$b$ to a linear function,
\begin{equation} \label{eq:ratio_tanb}
\frac{\sii}{\Ha} = \left(\frac{\sii}{\Ha}\right)_0 + \frac{\HHa}{\Hsii} \tan |b|.
\end{equation}
The best fit parameters are $(\sii/\Ha)_0 = 0.074 (0.093)$ and $\HHa/\Hsii = 2.05(1.79)$ above (below) the plane, with $\chi_r^2 = 1.94(0.77)$. Unlike the $\sii/\Ha$ versus \IHa, this fit is not appreciably affected by extinction. 

Is the observed behavior of the $\sii/\Ha$ line ratio primarily a function of height or \Ha\ intensity? To test this, we divided the line ratio by equation~(\ref{eq:ratio_tanb}) and plot the result as a function of $\IHac$ in Figure~\ref{fig:ratio}$c$, leaving $\sii/\Ha$ as a function of \Ha\ intensity with the height dependence removed. Similarly, we divided the line ratio by equation~(\ref{eq:ratio}) and plot the result as a function of $\tan b$ in Figure~\ref{fig:ratio}$d$, leaving $\sii/\Ha$ as a function of $z$ with the intensity dependence removed. Panel $c$ shows that a trend in $\sii/\Ha$ as a function of \IHac\ remains after this procedure, while panel $d$ shows little residual trend in $\sii/\Ha$ as a function of $\tan b$.

We tested the validity of our identified correlation using a Spearman rank correlation test, which tests for a monotonic relationship between two variables. We find a correlation coefficient of $r = -0.15$ between $\sii/\Ha$ with the $\tan b$ dependence removed and the \Ha\ intensity; with $N=1060$ data points, there is a $p \sim 10^{-6}$ probability that two uncorrelated data sets (for which we would have $r=0$) would produce $|r|$ this large, so we reject the null hypothesis that there is no correlation between these parameters. For $\sii/\Ha$ with the \Ha\ intensity dependence removed as a function of $\tan |b|$, we find $r = 0.022$. There is a $p \sim 0.5$ probability that an uncorrelated data set would produce $|r|$ this large, so we are unable to reject the null hypothesis that no residual trend remains in this case.

In addition, we tested for the significance of the correlation between two variables in the presence of a third variable using a partial Spearman correlation test\citep[see][]{Collins:1998jz,Urquhart:2013kd}. We use
\begin{equation}
r_{AB,C} = \frac{r_{AB} - r_{AC} r_{BC}}{\left[ \left(1-r_{AC}^2 \right) \left(1-r_{BC}^2 \right) \right]^{1/2}}
\end{equation}
for Spearman rank coefficients $r$ for each pair of parameters. We tested for a lack of correlation between $\sii/\Ha$ ($A$ in all cases) and $\IHa$ ($B$ here), removing their mutual dependence on $\tan |b|$ ($C$ here), finding $r_{AB,C} = -0.54$. We also tested for a correlation between $\sii/\Ha$ and $\tan |b|$ ($B$ here), removing their mutual dependence on $\IHa$ ($C$ here), finding $r_{AB,C} = -0.12$. In these cases, the $t$-values are $t \equiv r_{AB,C} [(N-3) / (1-r_{AB,C}^2)]^{1/2} = -20.8$ and $-4.1$, respectively. Both $t$ values are sufficient to reject the null hypothesis of no correlation between the variables, assuming $t$ is distributed according to a Student's $t$ distribution. However, These tests further support our conclusion that $\sii/\Ha$ is more strongly correlated with \IHa\ than with $\tan |b|$.

We thus conclude that the observed decrease in $\sii/\Ha$ in the Scutum-Centaurus Arm is primarily a function of \Ha\ intensity; the dependence of $\sii/\Ha$ on $|z|$ appears to be primarily a consequence of the typically-lower \Ha\ intensity at large $|z|$.

\section{Discussion and Conclusions} \label{sec:discussion}

The power-law relationship between $\sii/\Ha$ and \IHa, which is observed here as well as in local gas and the Perseus Arm, suggests that a relatively uniform physical process is at play across the observed range of \Ha\ intensities. The most likely explanation is a temperature effect, with higher $\sii/\Ha$ values corresponding to higher temperatures. A correlation between the $\nii/\Ha$ line ratio and the temperature of the emitting gas is now well established \citep{Haffner:1999hi,Collins:2001dz,Madsen:2006fw}. Ascribing the variation of $\sii/\Ha$ to temperature is less straightforward. The line ratio depends upon the temperature, sulfur abundance, and ionization state of the gas \citep{Otte:2002dr}:
\begin{equation}
\frac{\sii}{\Ha} = 7.49 \times 10^5 T_4^{0.4} \, e^{-2.14/T_4} \frac{\mathrm{H}}{\mathrm{H}^+} \frac{\mathrm{S}}{\mathrm{H}} \frac{\mathrm{S^+}}{\mathrm{S}}.
\end{equation}
The hydrogen ionization faction $\mathrm{H}^+ / \mathrm{H} \gtrsim 0.9$ in the WIM \citep{Reynolds:1998ji}, while we assume that the sulfur abundance $\mathrm{S}/\mathrm{H}$ is does not change appreciably in the diffuse ISM in the arm, leaving a dependence on the sulfur ionization fraction $\mathrm{S}^+/\mathrm{S}$ and the temperature.
Sulfur can be in either the $\mathrm{S}^+$ or the $\mathrm{S}^{++}$ state in photoionized gas because its first ionization potential is less than that of hydrogen, while the second ionization potential is $0.9 \textrm{ eV}$ below the helium edge. Photoionization modelling of sulfur is difficult because of the unknown temperature dependence of its dielectronic recombination rate \citep{Ali:1991ba,Haffner:2009ev,Barnes:2014wc}. However, because the observed variations in $\sii/\Ha$ largely track variations in $\nii/\Ha$ \citep{Haffner:1999hi,Madsen:2006fw,Haffner:2009ev}, it is likely that they are primarily tracing temperature variations, while changes in $\sii/\Ha$ relative to $\nii/\Ha$ (or, equivalently, changes in $\sii/\nii$) trace changes in the sulfur ionization state.

Our conclusion that $\sii/\Ha$ and, thus, the temperature of the gas depends primarily on \Ha\ intensity, not height, provides a clue about the non-photoionization heating of the WIM. The lower \Ha\ intensity presumably corresponds to a lower electron density in the emitting gas; if the lower intensity instead corresponded to a shorter path length, \sii\ would scale in the same way, leaving $\sii/\Ha$ unchanged. The temperature of the WIM is determined by the balance between heating and cooling, often parameterized as \citep{Reynolds:1999jp,Wiener:2013fl}
\begin{equation}
\Lambda n_e^2 = G_0 n_e^2 + G_1 n_e + G_2 + G_3 n_e^{-1/2},
\end{equation}
where $\Lambda$ is the cooling function. Pure photoionization heating ($G_1 = G_2 = G_3 = 0$) cannot explain the observed $\nii/\Ha$ line ratio of the WIM at low \Ha\ intensities: an additional heating source parameterized by non-zero $G_1$ (such as dissipation of turbulence or photoelectric heating of dust grains), $G_2$ (magnetic reconnection), or $G_3$ (cosmic ray heating) is required \citep{Reynolds:1992ci,Reynolds:1999jp,Otte:2002dr,Wiener:2013fl,Barnes:2014wc}. Assuming such a heating source is important, our conclusion that the $\sii/\Ha$ depends more strongly on \IHa\ than on $|z|$ suggests that the heating mechanism at a given $n_e$ does not vary significantly with $|z|$. This argues against photoelectric heating because the FUV radiation field has a significantly smaller scale height ($\sim 300 \pc$) than $n_e$. Cosmic rays, on the other hand, most likely have a larger scale height than the gas, so $G_3$ varies more slowly with height than $n_e$. However, more detailed modelling and \nii\ data to measure the temperature independent of the sulfur ionization state are required to test this.

The power-law relationship continues to $\IHac \approx 100 \R$ and then flattens. This change in slope suggests a physical change between the WIM and \ion{H}{2} regions. The $62$ sightlines with $\IHac > 100 \R$ regions have a mean $\langle \sii/\Ha \rangle = 0.082$ and a standard deviation $\sigma_{\sii/\Ha} = 0.037$, consistent with \ion{H}{2} region observations in other contexts. Alternatively, systematic uncertainty introduced by our extinction correction could produce this effect if the extinction to \ion{H}{1} column density ratio has increased scatter at large \ion{H}{1} columns. A spectroscopic survey of \Hb\ in the region we have mapped in \Ha\ would provide a more rigorous extinction correction.

The scale height and midplane densities we observe in the Scutum-Centaurus Arm at Galactocentric radius $r_G \approx 6 \kpc$ are intermediate between those observed in the inner Galaxy at $r_G \approx 4-6 \kpc$ \citep{Madsen:2005ft}, locally, and in the Perseus Arm at $r_G \approx 10 \kpc$ \citep{Haffner:1999hi,Madsen:2006fw}. In the inner Galaxy, the scale height is $\HHa \sim 95-150 \pc$, while we find $\HHa \approx 300 \pc$ in Scutum-Centaurus, and $\HHa = 400 \pc$ in Perseus. The corresponding space-averaged rms midplane electron densities are $f_0^{1/2} n_{e,0} = 0.5-1 \cucm$, $0.2 \cucm$, and $0.11 \cucm$, using the same assumptions as \citet{Madsen:2005ft}.

Our conclusion that the observed $\sii/\Ha$ line ratio depends primarily on the \Ha\ intensity of the sightline is relevant to recent claims that a significant fraction of the \Ha\ emission we attribute to the WIM in fact originated in higher-density regions and has been scattered into the beam by dust. \citet{Seon:2012ci} produced a model which they claim reproduces the high optical line ratios observed in the WIM with a large scattered light contribution. In their model, the diffuse \Ha\ emission primarily originates in late O and early B star \ion{H}{2} regions and is scattered by dust into high-latitude sightlines. O and B star \ion{H}{2} regions have typical line ratios $\sii/\Ha \le 0.15$ \citep{Reynolds:1988gp,Wood:2005cc,Madsen:2006fw}, inconsistent with the typical values of $\approx 0.4$ observed for the WIM either in local gas at high latitudes or in the Scutum-Centaurus or Perseus Arms. \citet{Seon:2012ci} argue that, in their picture, line ratios could increase with distance from the late O and early B \ion{H}{2} regions. However, our conclusion that the line ratio depends primarily on the electron density, not $|z|$ (Figure~\ref{fig:ratio}$c$ and $d$), qualitatively supports the interpretation that the variations in $\sii/\Ha$ trace temperature or density variations, not scattered light: we do not expect gas density variations at a fixed $|z|$ in the emitting regions of the WIM to correlate with distance from OB stars.

\section{Summary} \label{sec:summary}

We have investigated the properties of warm ($\approx 8000 \K$) ionized gas in the Scutum-Centaurus Arm $\approx 6 \kpc$ from the Galactic Center. The \Ha\ emission suggests a higher midplane density and lower electron scale height in the Scutum-Centaurus Arm than is observed near the Sun or in the Perseus Arm in the outer Galaxy. The higher -- though uncertain in the arm -- $n_e$ scale height can be reconciled if the volume filling fraction increases with height. The $\sii/\Ha$ line ratio, which is most likely a diagnostic of temperature, follows an inverse power law relationship with the \Ha\ intensity and an inverse linear relationship with height above the plane; the inverse correlation wtih intensity is stronger than with height above the plane. We interpret these observations as further evidence that the diffuse \Ha\ emission is primarily ($\gtrsim 80 \%$) {\em in situ} emission from photoionized gas far from ionizing stars. Scattered light may contribute a small fraction of the observed \Ha\ intensity but is a secondary effect.

\acknowledgements

We acknowledge useful discussions about photoionization with J.\ E.\ Barnes and K.\ A.\ Wood and Galactic structure with J.\ L.\ Caswell, J.\ A.\ Green, and N.\ M.\ McClure-Griffiths. We also thank G.\ J.\ Madsen for detailed comments on the manuscript and the anonymous referee for suggesting the statistical analysis in Section~\ref{sec:lineratios}. N.\ Chopra, N.\ Pingel and J.\ Wunderlin contributed to the data reduction. We thank K.\ Jaehnig, G.\ J.\ Madsen, and E.\ Mierkiewicz for their work in installing and calibrating WHAM at Cerro Tololo. WHAM is supported by the National Science Foundation through grant AST-1108911. KAB is supported through NSF Astronomy and Astrophysics Postdoctoral Fellowship award AST-1203059.

\bibliography{papers_bibtex}

\end{document}